# Efficient Spin-Orbit Torques in an Antiferromagnetic Insulator with Tilted Easy Plane

Pengxiang Zhang[1], Chung-Tao Chou[1,2], Hwanhui Yun[3], Brooke C. McGoldrick[1], Justin T. Hou[1], K. Andre Mkhoyan[3], and Luqiao Liu[1]

[1]*Department of Electrical Engineering and Computer Science, Massachusetts Institute of Technology, Cambridge, MA 02139, USA*

[2]*Department of Physics, Massachusetts Institute of Technology, Cambridge, MA 02139, USA*

[3]*Department of Chemical Engineering and Materials Science, University of Minnesota, Minneapolis, MN 55455, USA*

**Abstract**

**Electrical manipulation of spin textures inside antiferromagnets represents a new opportunity for developing spintronics with superior speed and high device density. Injecting spin currents into antiferromagnets and realizing efficient spin-orbit-torque-induced switching is however still challenging due to the complicated interactions from different sublattices. Meanwhile, because of the diminishing magnetic susceptibility, the nature and the magnitude of current-induced magnetic dynamics remain poorly characterized in antiferromagnets, whereas spurious effects further complicate experimental interpretations. In this work, by growing a thin film antiferromagnetic insulator, *α*-$Fe_2O_3$, along its non-basal plane orientation, we realize a configuration where an injected spin current can robustly rotate the Néel vector within the tilted easy plane, with an efficiency comparable to that of classical ferromagnets. The spin-orbit torque effect stands out among other competing mechanisms and leads to clear switching dynamics. Thanks to this new mechanism, in contrast to the usually employed orthogonal switching geometry, we achieve bipolar antiferromagnetic switching by applying positive and negative currents along the same**

**channel, a geometry that is more practical for device applications. By enabling efficient spin-orbit torque control on the antiferromagnetic ordering, the tilted easy plane geometry introduces a new platform for quantitatively understanding switching and oscillation dynamics in antiferromagnets.**

Spin-orbit torque (SOT) switching of antiferromagnets has been extensively pursued recently with both antiferromagnet single layers owning staggered spin torques[1-6] and antiferromagnet/heavy metal bilayer heterostructures[7-12]. In the latter case, it is expected that the spin Hall effect (SHE) from the neighboring heavy metal layer can act on the antiferromagnetic ordering and lead to Néel vector reorientation. Particularly antiferromagnets with easy-plane anisotropy like NiO, CoO and $\alpha$-Fe$_2$O$_3$ have been utilized in these experiments for achieving multiple equilibrium positions of the Néel vector. Since the magnetic easy plane is usually also the crystalline plane with low surface energy, antiferromagnetic films are almost always synthesized with their surface coinciding with the magnetic easy plane, the geometry of which unfortunately poses extra difficulties for controlling magnetic ordering with SOT. As shown in Fig. 1(a), since spins generated from standard SHE[13,14] are oriented in-plane at the antiferromagnet/heavy metal interface, the resulted damping-like torque $\tau_{DL}$, if any, tends to rotate the Néel vector out of the film plane. In this configuration, one needs to overcome the very strong easy-plane anisotropy in order to realize precession or switching[15,16], resulting in formidable threshold currents. Spurious thermal effects such as electromigration and magnetoelastic effects also emerge due to the large applied current[11,17-21], further shadowing real SOT-related physics.

Efficient control over the Néel vector can be potentially achieved if the injected spins form a finite angle with the magnetic easy plane. As shown in Fig. 1(b), when the magnetic easy plane is no longer parallel with the film surface, the SOT from the injected in-plane spins will have a component facilitating the Néel vector rotation within the easy plane. In this configuration, the SOT only needs to overcome the much weaker anisotropy within the easy plane. Under a $\tau_{\text{DL}}$, the effective fields on the normalized magnetic moments of the two sublattices $\boldsymbol{m}_{A(B)}$ have the form $\boldsymbol{H}_{\text{DL}}^{A(B)} \propto \boldsymbol{m}_{A(B)} \times \boldsymbol{\sigma}$, which rotate the two sublattices constructively, in contrast to the effect of an external magnetic field, which cancels between the two sublattices. Therefore, a threshold current similar to, or even smaller than that in traditional ferromagnet can be achieved in antiferromagnets with this tilted easy plane. In this work, we realized the SOT configuration in Fig. 1(b) by growing antiferromagnetic thin film α-Fe$_2$O$_3$ along its *R*-plane, a non-basal plane orientation. Different from previously studied *C*-plane samples where the SOT effect remains hardly detectable[18], spins injected from an adjacent platinum film lead to very efficient Néel vector rotation in this *R*-plane sample, which was further quantitatively calibrated with real magnetic fields. Utilizing SOT, we also achieved bipolar switching by applying positive and negative currents along the same path, in contrast to previously studied geometries utilizing two orthogonal current paths[1,4,8]. Besides magnetic switching, configurations demonstrated in our experiment can also be utilized for realizing low power antiferromagnetic oscillator[22], as well as magnon spin superfluidity[23-25] that have been predicted in previous literatures.

α-Fe$_2$O$_3$ is a well-studied antiferromagnetic insulator with high Néel point (955 K) and easy-plane anisotropy at room temperature[26,27]. Because of the very weak magnetic anisotropy within the easy plane (*C*-plane)[28], magnetic field needed for spin-flop transition is unusually low in this antiferromagnet (<1 Tesla), enabling people to control the Néel vector easily. Besides the common *C*-plane (0001) orientation, α-Fe$_2$O$_3$ thin film has also been grown along a few other low-index directions, including *A*- ($\bar{2}110$) and *R*-plane ($01\bar{1}2$) [Fig. 2(a)], both of which satisfy the finite angle requirement in Fig. 1(b). Meanwhile, in order to monitor the Néel vector through the spin Hall magnetoresistance (SMR), an oblique, rather than right angle between the easy-plane and the sample surface is preferred[29]. Therefore, we choose to focus on $R$-plane films in this work, which are epitaxially grown on $R$-plane $\alpha$-Al$_2$O$_3$ substrates. As the $R$-plane becomes the horizontal film surface, the easy plane ($C$-plane) forms a tilting angle $\chi_{\text{tilt}} = 58°$ with the surface [Fig. 2(a)]. To reduce the strain from the lattice mismatch between $\alpha$-Fe$_2$O$_3$ and $\alpha$-Al$_2$O$_3$ (~5.8%), we deposit 1 nm of $\alpha$-Cr$_2$O$_3$ prior to $\alpha$-Fe$_2$O$_3$ as the seeding layer, which has an intermediate lattice constant. From magnetometry and electrical measurement as discussed below, we found that the 1 nm $\alpha$-Cr$_2$O$_3$ behaves inactively, making no observable contribution to magnetic dynamics. The crystal structure of the $\alpha$-Fe$_2$O$_3$ film is examined using X-ray diffraction (XRD) [Fig. 2(c)], where the $(01\bar{1}2)$ diffraction peak position agrees with the expected lattice constant and the film growth direction. The epitaxial relationship between the deposited film and the substrate is further verified through the XRD reciprocal space mapping (RSM) [Fig. 2(d)]. The RSM, together with the scanning transmission electron microscope (STEM) image [Fig. 2(b)] shows that the strain is mostly relaxed at the film/substrate interface via misfit dislocations (See Supplemental Information for details).

We characterize magnetic properties of the films using superconducting quantum interference device (SQUID) magnetometer at 300 K. When a field $H$ is applied in the film plane along the cleavage edge of the sample, we see a typical $M$-$H$ loop from canted antiferromagnetism [Fig. 2(e)]. It is known that in its easy-plane phase, the sublattices in $\alpha$-Fe$_2$O$_3$ form a very small canting angle ($< 0.1°$) through the Dzyaloshinskii-Moriya (DM) interaction[28], which induces a tiny net magnetization (1~2 emu/cm$^3$) and provides a handle for controlling the Néel vector. The measured small, finite magnetization therefore suggests that the $R$-plane sample remains in the easy-plane phase, similar to $C$-plane ones studied earlier. But different from $C$-plane samples with low coercivity ($H_C \sim 1$ kOe), we find $H_C$ in the $R$-plane sample is larger (10 kOe). Moreover, by applying field along the $z$ axis, we also measure the out-of-plane $M$-$H$ loop [Fig. 2(f)], where a spin-flop transition is observed, indicating additional, high order magnetic anisotropy developed on top of the standard easy-plane anisotropy. With $M$-$H$ curves measured along different directions as well as electrical magnetoresistance measurements under rotating fields (discussed below), we conclude that a weak, uniaxial magnetic anisotropy exists within the $C$-plane, with the easy axis along the $[0\bar{1}10]$ direction, or the $a$-axis defined Fig. 2(a). This additional anisotropy within the easy plane is likely caused by the growth-induced symmetry breaking, where the residual strain distorts the hexagonal $C$-plane along the $a$-axis, making it the preferred axis through magnetoelastic effect. Here we note that despite of the increased in-plane $H_C$, it does not prevent us from observing SOT's effects since $\tau_{DL}$ acts constructively on the Néel vector [Fig. 1(b)] while the measured $H_C$ reflects field's effect which largely cancels between sublattices.

To study the SOT, we sputter 5 nm Pt on $R$-plane $\alpha$-Fe$_2$O$_3$ of 30 nm thick and fabricate Hall bars of 8 μm wide with the current channel aligned along the intersection line between $R$- and $C$-plane [Fig. 3(a)]. We measure the transverse SMR at room temperature while applying a rotational magnetic field $\boldsymbol{H}$ within the $xz$-plane, with the field angle $\beta$ defined in Fig. 3(a). When the projected component of $\boldsymbol{H}$ on the $C$-plane is larger than the spin-flop field, it will align the net moment $\boldsymbol{m}_{net} = (\boldsymbol{m}_A + \boldsymbol{m}_B)/2$ parallel with, and the Néel vector $\boldsymbol{n} = (\boldsymbol{m}_A - \boldsymbol{m}_B)/2$ perpendicular to it. In the device shown in Fig. 3(a), when a current flows along the $x$-axis, spins $\boldsymbol{\sigma}$ polarized along $y$ from the SHE lead to antiferromagnetic SMR signal $R_H^{SMR} = -R_0^{SMR} n_x n_y = -\frac{1}{2} R_0^{SMR} \sin 2\varphi \cos \chi_{tilt}$,[29] where $R_0^{SMR}$ is the transverse SMR amplitude, and $n_x = -\sin\varphi$ and $n_y = -\cos\varphi \cos\chi_{tilt}$ are components of $\boldsymbol{n}$, with $\varphi$ being the azimuthal angle of $\boldsymbol{n}$ in the $C$-plane [see Fig. 3(a)]. Fig. 3(b) shows the measured $R_H^{SMR}$ (or $R_H^\omega$ for a first harmonic lock-in measurement) as a function of $\beta$ under different field strengths, after an ordinary Hall resistance is subtracted. The ordinary Hall resistance from Pt linearly depends on the z component of $\boldsymbol{H}$ and has a 360° periodicity as a function of $\beta$, which can be calibrated in a standard way (Details in Supplemental Information). In Fig. 3(b), we find that $R_H^\omega$ has an angular dependence with 180° period, consistent with the $\sin(2\varphi)$ factor in its formula. Under lower magnetic fields ($H < 12$ kOe), $R_H^\omega$ exhibits magnetic hysteresis, while for $H > 20$ kOe, $R_H^\omega$ is smoother and agrees better with a sinusoidal function, in consistency with the switching fields measured in Fig. 2(e) and 2(f). Moreover, in Fig. 3(b) we notice that the slopes for $-45° < \beta < 45°$ are flatter compared with the ones for $45° < \beta < 135°$, in agreement with the fact that $a$ axis is the easy axis within the easy plane. Besides the $\sin(2\beta)$ dependence, we find that $R_H^\omega$ in Fig. 3(b) has a small, residual component with a 360° periodicity, which maximizes (minimizes) for $\boldsymbol{H}$ along the $+z$ ($-z$) direction. This anomalous-Hall-like signal may originate from magnetic proximity or crystal Hall effect[30], both of which have the symmetry of $R_H \propto m_z^{net}$. As this signal is relatively small and can be separated from the main SMR via its angular dependence, we choose not to expand on its root origin.

The equilibrium orientation of $\boldsymbol{n}$, hence $R_H^{SMR}$ under an applied $\boldsymbol{H}$ can be determined by considering the magnetic free energy $F = E_Z + E_{ex} + E_{DM} + E_{an}$. Here, the Zeeman energy, exchange energy, energy due to DM interaction and anisotropy energy are written as $E_Z = -\mu_0 M_0 (\boldsymbol{m_A} + \boldsymbol{m_B}) \cdot \boldsymbol{H}$, $E_{ex} = \mu_0 M_0 H_{ex} \boldsymbol{m_A} \cdot \boldsymbol{m_B}$, $E_{DM} = -\mu_0 M_0 H_{DM} \hat{\boldsymbol{c}} \cdot (\boldsymbol{m_A} \times \boldsymbol{m_B})$, and $E_{an} = K_1[(\boldsymbol{m_A} \cdot \hat{\boldsymbol{c}})^2 + (\boldsymbol{m_B} \cdot \hat{\boldsymbol{c}})^2] - K_2[(\boldsymbol{m_A} \cdot \hat{\boldsymbol{a}})^2 + (\boldsymbol{m_B} \cdot \hat{\boldsymbol{a}})^2]$, where $\mu_0$ and $M_0$ are the vacuum permeability and the saturation magnetization of one sublattice, $H_{ex}$ and $H_{DM}$ are effective fields from the exchange and the DM interaction, $K_1$ and $K_2$ are the energy densities for easy-plane and easy-axis anisotropy. Because of the very strong easy-plane anisotropy, $\boldsymbol{m_A}$ and $\boldsymbol{m_B}$ are mostly confined within the C-plane in our experiment. As an approximation, we only consider their degrees of freedom within the C-plane, quantified by the two angles $\varphi$ and $\delta$ defined in Fig. 3(a). For a given $\boldsymbol{H}$, we can determine the equilibrium angle $\varphi_0$, $\delta_0$ through $\frac{\partial F(\varphi,\delta)}{\partial \varphi} = \frac{\partial F(\varphi,\delta)}{\partial \delta} = 0$ (Supplemental Information). $R_H^{SMR}$ is further calculated with the obtained $\varphi_0$, as shown by the solid black lines for $H = 15$ kOe - 23 kOe, in Fig. 3(b), which have good agreement with the experimental data. Using material parameters reported in literatures[26,31,32]: $M_0 = 759$ emu/cm$^3$, $H_{ex} = 9000$ kOe, $H_{DM} = 17.8$ kOe, and $K_1 = 7.6 \times 10^4$ erg/cm$^3$, we determine the single fitting parameter $K_2 = 4.9 \times 10^3$ erg/cm$^3$, which universally fits all of the experimental curves in Fig. 3(b).

We quantify the SOT by detecting the current-induced Néel vector rotation. As shown in Fig. 3(a), in the presence of SOT, $\boldsymbol{n}$ undergoes an additional rotation $\Delta\varphi$ from its equilibrium. When $\Delta\varphi$ is caused by an alternating current (a.c.), a change in $R_H^{SMR}$ with the same frequency appears, therefore a voltage at the second harmonic frequency appears ($V_H^{2\omega}$), which can be used to determine the SOT, as has been widely used in studies on ferromagnets[33]. Fig. 3(c) shows the second harmonic resistance $R_H^{2\omega} = V_H^{2\omega}/I$. Outside the hysteresis region ($H \geq 15$ kOe), there are two peaks close to $\beta = \pm 90°$, and the peak magnitude decreases when $H$ increases. The position and the field magnitude dependence of $R_H^{2\omega}$ are consistent with the signature of SOT. Firstly, at $\beta = \pm 90°$, the slope of $\frac{dR_H^{SMR}}{d\beta}$ reaches maximum according to Fig. 3(b), which therefore converts the $\boldsymbol{n}$ rotation to $R_H^{2\omega}$ most sensitively. As for the field strength dependence, a higher $H$ pins $\boldsymbol{n}$ more strongly, and suppresses the current-induced rotation, resulting in a smaller $R_H^{2\omega}$. We also verify that $R_H^{2\omega}$ is proportional to the applied current [Fig. 3(e)], as expected from the SOT mechanism, while the first harmonic resistance shows no current dependence [Fig. 3(d)].

To quantify the SOT, we model $R_H^{2\omega}$ in an antiferromagnet. In general, $R_H^{2\omega}$ can be expressed as $R_H^{2\omega} = \frac{1}{2}\frac{dR_H^{\omega}}{d\varphi}\Big|_{\varphi_0} \cdot \Delta\varphi(I_0)$, where $\frac{dR_H^{\omega}}{d\varphi}\Big|_{\varphi_0}$ is directly derived from SMR's formula, and the tilting angle at peak current $\Delta\varphi(I_0)$ is determined through the balance condition between the SOT and the torque due to the magnetic free energy $\tau_{ST}(I_0) + \tau_M(I_0) = 0$. Under the approximation that the current-induced $\delta$ angle change is negligible due to the very strong exchange field, $R_H^{2\omega}$ can be calculated as (Supplemental Information):

$$R_H^{2\omega} \approx R_0^{SMR} \cos\chi_{tilt} \cos 2\varphi_0 \frac{H_{DL}(I_0)\sin\chi_{tilt} + H_{FL}(I_0)\cos\chi_{tilt}\cos\frac{\delta_0}{2}\cos\varphi_0}{(4K_2\cos 2\varphi_0 \cos\delta_0)/\mu_0 M_0 - 2H\cos\frac{\delta_0}{2}(\cos\beta\cos\varphi_0 + \sin\chi_{tilt}\sin\beta\sin\varphi_0)}, \quad (1)$$

where $H_{DL}(I_0)$ and $H_{FL}(I_0)$ are the damping-like and field-like effective fields on individual sublattices as in Fig. 3(a). With SOT theory, $H_{DL}(I_0)$ is further calculated to be $H_{DL} = \frac{1}{2}\frac{\hbar \xi_{DL} J_C}{2e\mu_0 M_0 t}$,[34] where $\hbar, e, \mu_0, J_C, t$ and $\xi_{DL}$ are the reduced Planck constant, electron charge, permeability of vacuum, peak current density in Pt, $\alpha$-Fe$_2$O$_3$ thickness, and the efficiency of $\tau_{DL}$. Here we neglect the very small contribution from current-induced Oersted field, which has similar symmetry with $H_{FL}$. Eq. (1) shows that $H_{DL}$ and $H_{FL}$ own different angle dependences due to their opposite symmetries under ***n*** reversal, allowing independent determination of the two quantities. We fit the data in Fig. 3(c) using $R_H^{2\omega}$'s formula above with only $\xi_{DL}$ and $H_{FL}$ as fitting parameters, and the results are shown by the solid black lines, which are in good agreement with all the experimental curves. From the fitting, we obtain $\xi_{DL} = 0.015$ and $H_{FL} = 100$ Oe under a current of $1.4 \times 10^7 \text{A/cm}^2$. This $\xi_{DL}$ is smaller than the intrinsic spin Hall angle of Pt $(0.05\sim 0.3)$[34,35], but comparable to values reported in Pt/ferrimagnetic insulator [36,37], probably due to a lower interfacial spin-mixing conductance with insulating material. The similar magnitude in damping-like SOT between ferromagnet and antiferromagnet therefore proves that $\tau_{DL}$ can be an efficient mechanism for controlling the Néel vector in the titled easy-plane geometry.

We further verify the SOT origin of the measured signal with control experiments by flipping the current and voltage terminals, i.e., applying current along $y$ and measuring voltage along $x$ as defined in Fig. 3(a). Under this geometry, the injected spins lie within the $C$-plane and the damping-like torque causes an out-of-easy-plane, instead of in-easy-plane rotation, similar to Fig. 1(a). The negligible damping-like torque signal in our observation (see Supplementary Information) agrees with this picture. Meanwhile, we can also exclude other thermally induced effects such as magnetoelastic effect from our measurement. This is because for any thermal cause, the Néel vector tilting will depend linearly on the temperature variation, and quadratically on the current, which, when mixed with the current, gives rise to a 3rd harmonic rather than 2nd harmonic voltage.

While the harmonic measurement with a.c. provides a quantitative method to characterize the SOT, the rotation of $\mathbf{n}$ can also be more intuitively captured through a direct current (d.c.) measurement. As illustrated in Fig. 1(b), the role of $\tau_{DL}$ is to cause an almost constant rotation angle on $\mathbf{n}$ when $\beta$ is varied, therefore, this will result in a horizontal shift in the $R_H^{SMR} - \beta$ curve. This is demonstrated in Fig. 4(a) for $I = \pm 6$ mA, which has a systematic shift of $\Delta\beta$ between the two curves, with the shift direction agreeing with the sign of $\tau_{DL}$ determined from the harmonic measurement. Since we compare $\pm I$ with the same magnitude, artifacts with thermal origins get canceled between the two. We note that the waveform of d.c. SMR curves in Fig. 4(a) is not exactly the same with the a.c. results in Fig.3(b), and the latter has stronger hysteresis. This is because we conditioned the samples with a large current (11 mA) before the d.c. measurement, as previous studies show the current-induced annealing effect can change magnetic anisotropy and make the d.c. measurement more repeatable[11]. When the applied field is much larger than the anisotropy field, we can assume the current induced $\boldsymbol{n}$ rotation roughly follows the angle shift of the applied $H$, i.e., $\Delta\varphi \approx \Delta\beta$. In Fig. 4(b) we summarize $\Delta\beta$ measured under currents and fields, where $\Delta\beta$ scales proportionally with $I$ and inversely proportionally with $H$, in agreement with the expected rotation angle of $\Delta\varphi \approx \frac{H_{\mathrm{DL}}\sin\chi_{\mathrm{tilt}}}{H\cos\frac{\delta_0}{2}}$ (see Supplemental Information). Linear fittings on curves in Fig. 4(b) lead to $\xi_{\mathrm{DL}} = 0.018 \pm 0.003$, close to the second harmonic results.

Using the tilted easy plane geometry, we also observe a current-induced switching. Fig. 5(c) shows $R_{\mathrm{H}}^{\mathrm{SMR}}$ measured after applying positive and negative current pulses along the same channel. Under high current ($|I| \geq 13$ mA), finite differences in the remnant $R_{\mathrm{H}}^{\mathrm{SMR}}$ values develop after pulses with opposite polarities are applied. During this measurement, a constant $H$ close to the spin-flop field is applied along the $z$-axis, to compensate the in-plane anisotropy field and assist the current-induced switching[23]. We note that consistent with this picture, switchings only happen under fields with intermediate strength (4 kOe $\leq |H_z| \leq$ 6 kOe), while a too high or too low $H_z$ does not allow switching due to the large net effective field after the cancellation between $H_z$ and anisotropy field [Fig. 5(d)]. This field dependence points to a magnetic origin, since non-magnetic artifacts like electromigration should be insensitive to the field conditions. The current-induced switching can be understood through the schematic illustration in Fig. 5(a). When the applied $H_z$ becomes comparable to the in-plane anisotropy field, $\tau_{\mathrm{DL}}$ under positive (negative) current rotates $\boldsymbol{n}$ along the clockwise (counterclockwise) direction, which causes a difference in the SMR value at the remnant state. In Fig. 5(d), we also see that the polarity of switched $\Delta R_{\mathrm{H}}^{\mathrm{SMR}}$ remains the same when $H$ reverses from +5 kOe to –5 kOe, in agreement with the SOT mechanism. As illustrated in Fig. 5(a) and (b), the rotation direction of $\bm{n}$ under a pair of positive and negative currents remains the same $2\Delta\varphi(I)$, independent on the sign of $H$. Meanwhile, under the reversal of $H$, the equilibrium angle of $\bm{n}$ changes from $\varphi_0$ to $\varphi_0 + 180°$, yielding the same $\Delta R_H^{SMR} = -R_0^{SMR} \cos\chi_{tilt} \cos 2\varphi_0 \cdot 2\Delta\varphi(I)$.

The bipolar switching in Fig. 5 represents a new form of switching as currents only flow along a single channel. Previously in Pt/antiferromagnet structures with current pulses applied along two orthogonal channels, thermally induced magnetoelastic effect can play a dominant role due to the anisotropic heating[9,18,21]. In our case, such effects make no contribution as Joule heating remains the same under the reversal of $I$. The switching magnitude is small in our experiment, which corresponds to ~2% of the total SMR signal. This small portion may be related to the weak hysteresis in the magnetization loop of Fig. 2(f), where two remnant states under the same $H_z$ bear small differences. Future efforts on increasing the remanence can be useful to enhance the switching signal.

To conclude, we experimentally demonstrate current-induced magnetic rotation and switching in an antiferromagnetic insulator with a tilted easy plane. In our experiment, spins injected along an oblique angle with respect to the easy plane provide an efficient mechanism to reorient Néel vectors. Aside from this titled easy plane geometry, the finite angle can also be achieved by using SHE materials with reduced symmetry like WTe$_2$, Mn$_3$GaN and Mn$_3$Ir, where an out-of-film-surface spin component can be generated from the non-conventional SHE[38-41]. Our work provides a new platform where SOT stands out from spurious effects in antiferromagnetic switching. The concept proved in our experiment can also help to investigate a rich family of antiferromagnetic dynamics such as THz spin torque oscillation and spin superfluidity.

**Methods**

Our $\alpha$-Fe$_2$O$_3$ films were grown on $\alpha$-Al$_2$O$_3$ $(01\bar{1}2)$ [$R$-plane] substrates by rf magnetron sputtering [AJA Orion] of an $\alpha$-Fe$_2$O$_3$ target, at a base pressure of less than $3\times10^{-8}$ Torr, Ar pressure 2 mTorr, and substrate at room temperature. An $\alpha$-Cr$_2$O$_3$ seeding layer was sputtered before $\alpha$-Fe$_2$O$_3$ at the same condition, without breaking the vacuum. To overcome oxygen vacancies and improve crystal quality, we annealed our films in ambient pressure of 50% O$_2$ 50% N$_2$ gas at 700-900°C for 1 hour. The SQUID magnetometry [Quantum Design MPMS3] was used to characterize the magnetic properties of the obtained film. Due to the tiny magnetic moment, careful efforts have been made to avoid contamination and contributions from sample holder. Results from sample films were compared with baseline measurements of bare $\alpha$-Al$_2$O$_3$ substrates, which was used for excluding substrate effects. The dimensions of samples for in-plane SQUID measurements are $\sim 6 \text{ mm} \times 6 \text{ mm} \times 0.43 \text{ mm}$, to fit into the plastic straw sample holder. For out-of-plane measurements, the dimensions are $\sim 3 \text{ mm} \times 3 \text{ mm} \times 0.43 \text{ mm}$, and the sample was glued to a larger piece of bare $\alpha$-Al$_2$O$_3$ carrier substrate. The X-ray diffraction was taken from high-resolution XRD [Bruker D8], using Cu K$_{\alpha 1}$, with 0D [Pathfinder] and 1D [LynxEye] detectors. HAADF-STEM imaging was carried out using aberration-corrected FEI Titan G2 60-300 (S)TEM. The TEM operating parameters were beam energy of 200 keV, a beam current of 30 pA, probe convergent semi-angle of 25 mrad. Cross-sectional TEM sample was prepared by using focused ion beam (FIB) lift-out method using an FEI Helios Nanolab G4 dual-beam FIB.

We later sputtered 5 nm Pt on the 30 nm $\alpha$-Fe$_2$O$_3$ film and fabricated it into Hall bars with 8 μm width using direct-write photolithography [MLA 150] and ion-milling [IntlVac Nanoquest]. The sample die [$\sim 1.5 \text{ mm} \times 1.0 \text{ mm} \times 0.43 \text{ mm}$] was mounted onto a rotating electrical transport sample holder, which is driven by a stepper motor, and the Hall bar to measure was wire-bonded.

The Hall bar measured was located at the center of a dipole electromagnet, whose maximum magnetic field is 23 kOe, and field accuracy is better than 1%. Angle-dependent and field-dependent transverse resistances were measured. The a.c. measurements were taken by a lock-in amplifier [EG&G 7260]. The d.c. and pulse switching measurements were taken by a combination of a d.c. source meter [Keithley 2400] and a sensitive d.c. voltmeter [Keithley 182].

**Author contributions**

L.L. and P.Z. conceived the project. P.Z. synthesized the materials and characterized the structural and magnetic properties, with the help of C.-T.C. and B.C.M.. H.Y. and K.A.M. performed the electron microscopy measurements and analysis. P.Z. and J.T.H. fabricated the devices. P.Z. performed the electrical measurements with the help of C.-T.C., B.C.M., and J.T.H.. P.Z. and L.L. analyzed the data and constructed the theoretical model. P.Z. and L.L. wrote the manuscript. All authors discussed the results and commented on the manuscript.

**Competing financial interests**

The authors declare no competing financial interests.

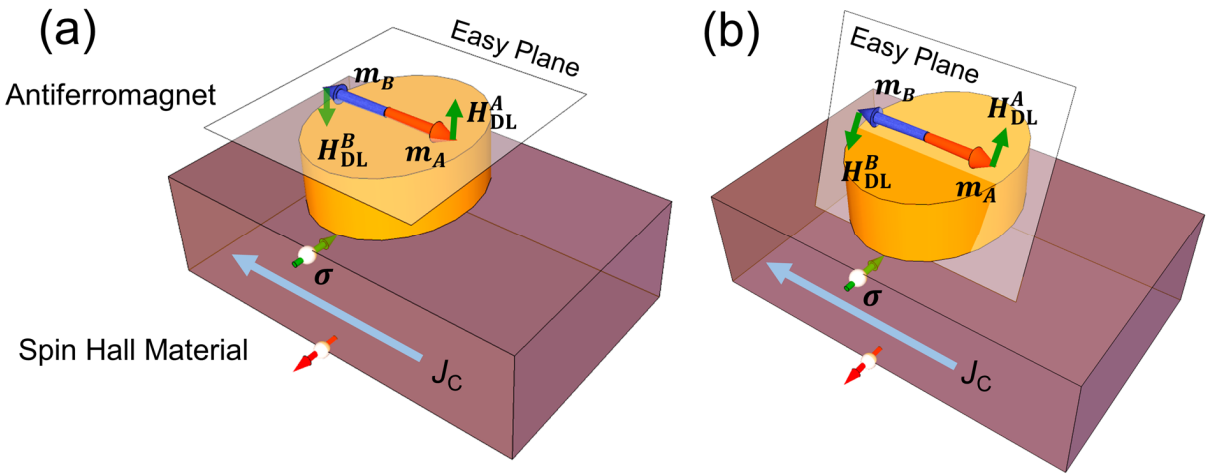

**Figure 1 Schematics of damping-like SOT on antiferromagnets with different easy plane orientations.** (a) Antiferromagnets with its easy plane being parallel with the film surface. The effective fields from damping-like torque $H_{DL}^{A(B)}$ try to rotate the two magnetic moments $m_{A(B)}$ towards out-of-plane direction. (b) Antiferromagnets with its easy plane forming a finite angle with the film surface. $H_{DL}^{A(B)}$ have components for rotating $m_{A(B)}$ within the easy plane. Only a very small energy barrier needs to be overcome in this geometry.

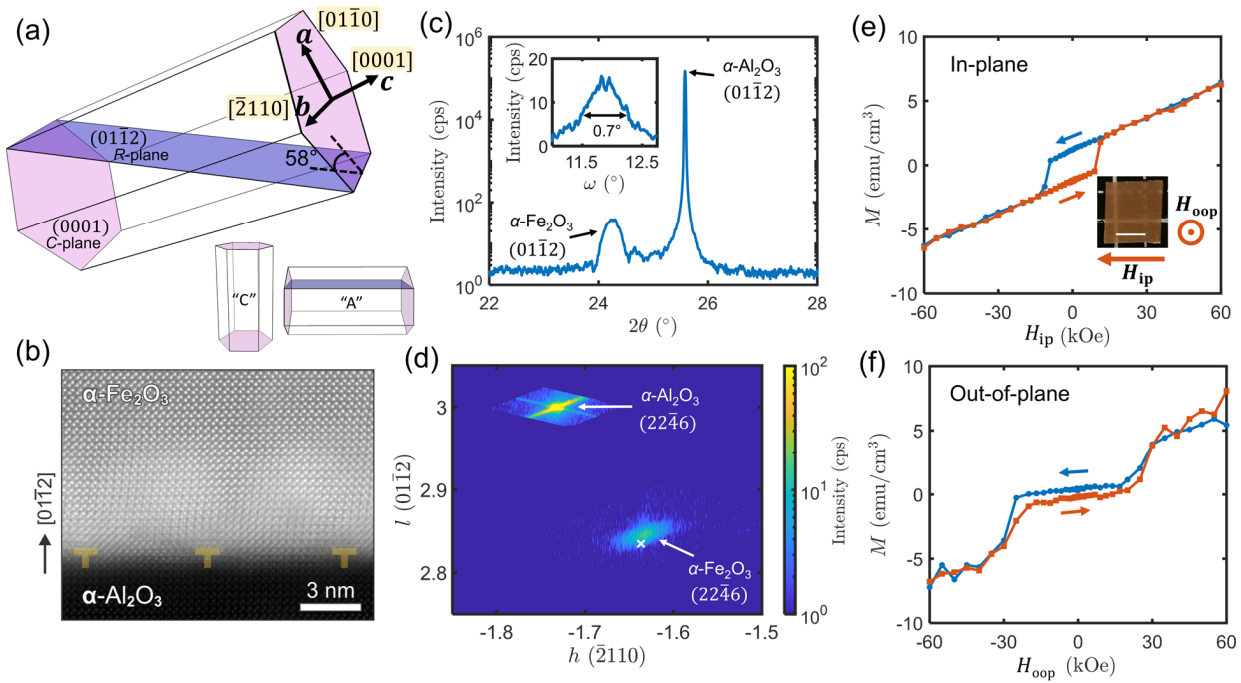

**Figure 2 Structural and magnetic properties of *R*-plane α-Fe₂O₃ thin film.** (a) Schematic of a unit cell in *R*-plane film. Inset: schematics of unit cell orientation for *C*-plane ("C") and *A*-plane ("A") samples. (b) Cross-sectional atomic-resolution high-angle annular dark-field (HAADF) STEM image of an *R*-plane sample. Dislocation symbols are marked. (c) Result from an XRD symmetric $2\theta$-$\omega$ Scan. The unit of 'cps' represents counts per second. Inset: Rocking curve of the XRD measurement. (d) XRD reciprocal space map, showing the $(22\bar{4}6)$ peak from the substrate (α-Al₂O₃) and the film (α-Fe₂O₃). The cross symbol represents the diffraction peak center position if the strain in the film were fully relaxed (i.e., with bulk lattice constant), which is very close to the experimentally obtained peak position. (e) and (f) *M-H* loop from SQUID magnetometry for in-plane (e) and out-of-plane (f) field application. Inset of Fig. (e): Photo of the *R*-plane sample for SQUID measurement, with the field direction labelled. The white scale bar is 3 mm.

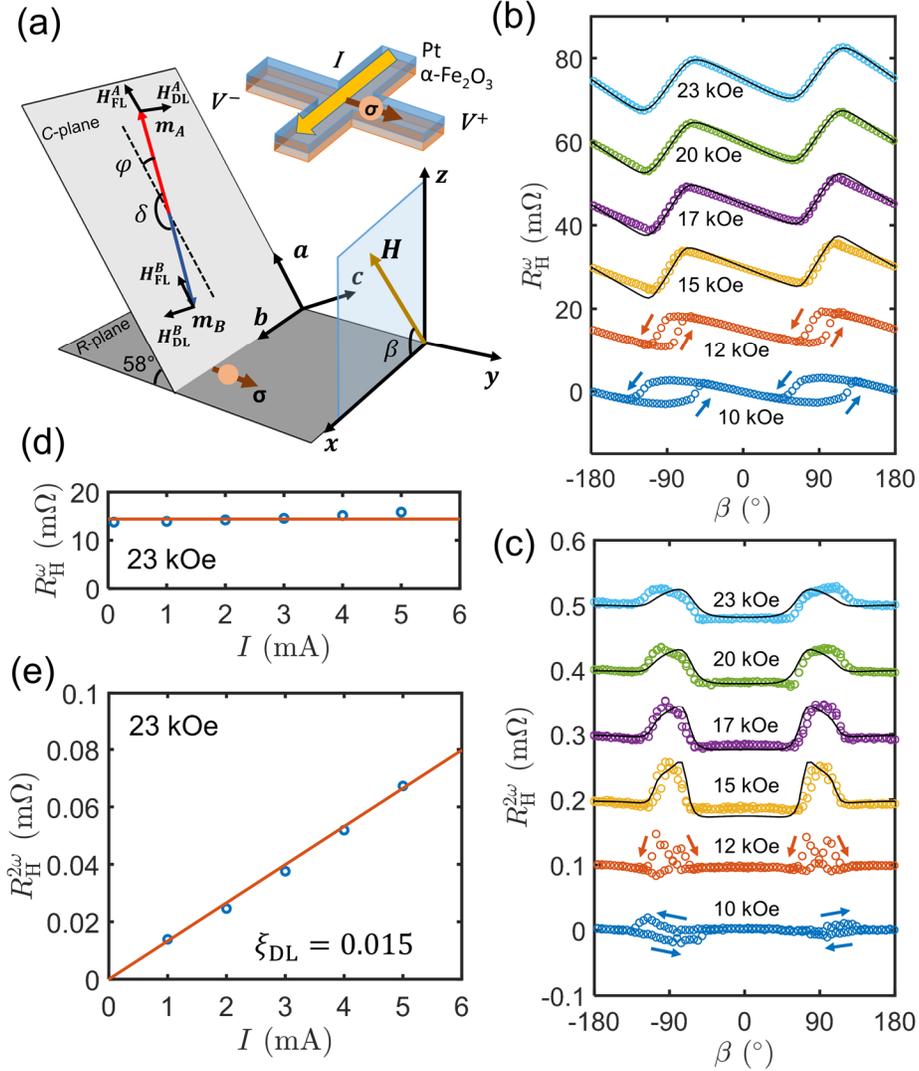

**Figure 3 Harmonic measurement on transverse SMR for SOT determination.** (a) Schematic of the SOT geometry in *R*-plane sample. Current is applied along the $x$ axis in the *R*-plane and the injected spins $\sigma$ are in the $y$ direction. A field with an angle of $\beta$ is applied within the $xz$ plane. $\varphi$ and $\delta$ define the angles of $n$, and the spanning angle between $m_{A(B)}$ within the easy plane, separately. Top right inset: schematic of the tested Hall bar device. (b) and (c) First (b) and second (c) harmonic results of the transverse SMR as a function of $\beta$. In those two figures, the root mean square value of the applied current is 4 mA. (d) and (e) Current dependence of the first and second harmonic transverse SMR. Results in both figures represent peak-to-peak values extracted from measurements at $H = 23$ kOe.

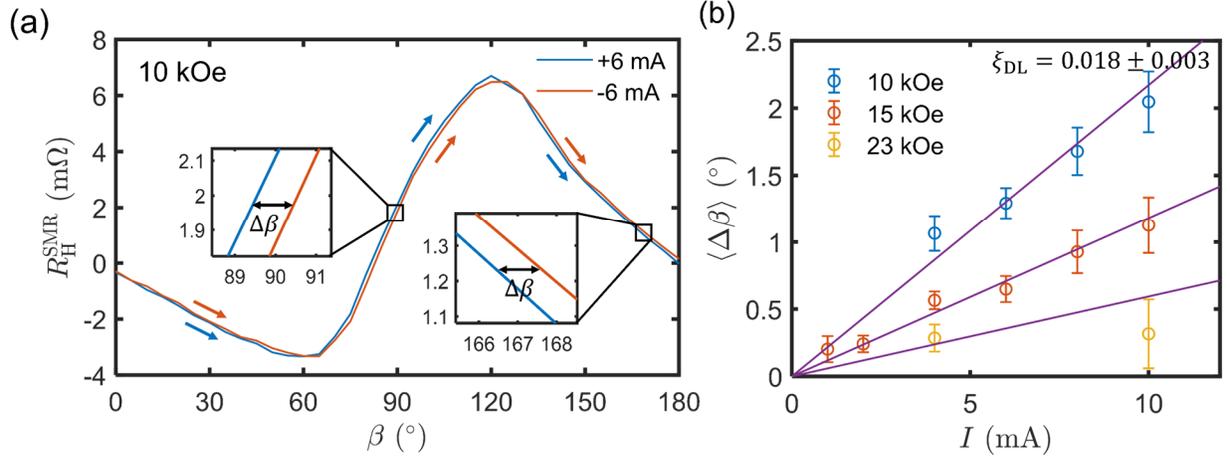

**Figure 4 d.c. measurement on SOT efficiency through current-induced Néel vector rotation.** (a) $R_H^{SMR}$ as a function of $\beta$ obtained with testing current of $I = \pm 6$ mA. The insets show magnified view of two typical regions on the curves to illustrate the current-induced horizontal shift angle $\Delta\beta$. (b) Summary of the current magnitude dependence of $\Delta\beta$ tested under different applied fields.

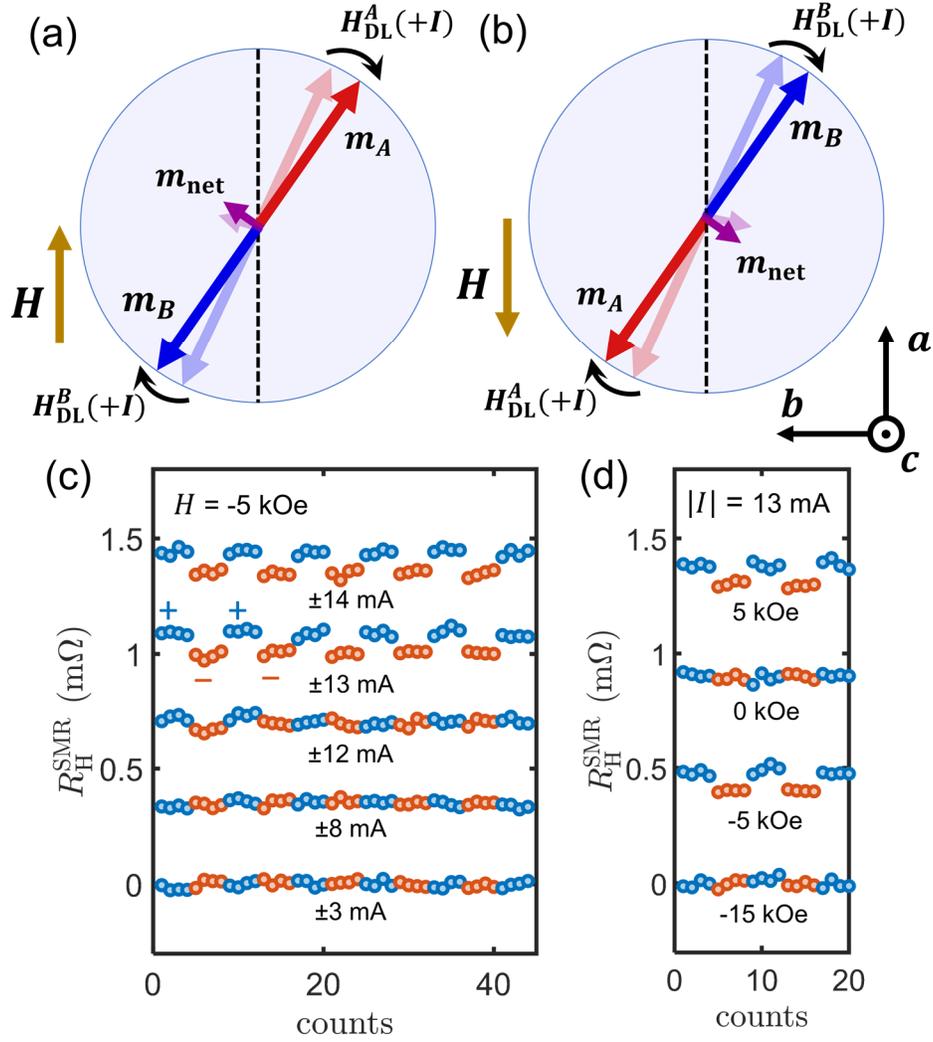

**Figure 5 Current induced bipolar switching.** (a) and (b) Schematics of the current-induced Néel vector rotation in the $C$-plane for positive (a) and negative (b) applied fields. The darker (lighter) arrows represent the position of $\boldsymbol{n}$ after positive (negative) current pulses. For $\boldsymbol{H}$, only the component projected onto the $C$-plane is shown here. (c) Switching of $R_H^{SMR}$ under current pulses with different magnitudes. The blue (red) dots show the measured resistances after positive (negative) current applications. (d) The dependence of $R_H^{SMR}$ switching on the applied field. Switching is observed for intermediate fields (5 kOe). The switching polarity remains the same between $\pm 5$ kOe, consistent with the schematics in (a) and (b).